\documentclass[fleqn,usenatbib]{mnras}

\usepackage[T1]{fontenc}
\usepackage{ae,aecompl}

\usepackage{graphicx}	% Including figure files
\usepackage{amsmath}	% Advanced maths commands
\usepackage{amssymb}	% Extra maths symbols
\usepackage{multicol}   % Multicolumns
\usepackage{bm}         % Vectors

\newcommand{\braket}[1]{\left\langle #1 \right\rangle}

\title[Cosmological Alignment of Radio Sources]{Radio Galaxy Zoo: Cosmological Alignment of Radio Sources}

\author[O. Contigiani et al.]{O. Contigiani,$^{1}$\thanks{E-mail: contigiani@strw.leidenuniv.nl}
F. de Gasperin,$^{1}$
G. K. Miley,$^{1}$ 
L. Rudnick$^{2}$,
H. Andernach,$^{3}$ \newauthor
J. K. Banfield,$^{4, 6}$ 
A. D. Kapi\'nska,$^{5, 6}$
S. S. Shabala$^{7}$,
O. I. Wong$^{5, 6}$
\\
$^{1}$ Leiden  Observatory, Leiden  University,  P.O. Box 9513, 2300 RA,  Leiden,  the Netherlands
\\
$^{2}$ Minnesota Institute for Astrophysics, University of Minnesota,
116 Church St. SE, Minneapolis, MN 55455
\\
$^{3}$ Departamento de Astronom\'ia, DCNE, Universidad de Guanajuato, Apdo.
Postal 144, CP 36000, Guanajuato, Gto., Mexico
\\
$^{4}$ Research School of Astronomy and Astrophysics, Australian National University, Weston Creek, ACT 2611, Australia
\\
$^{5}$ International Centre for Radio Astronomy Research (ICRAR), The University of Western Australia, 
\\
M468, 35 Stirling Hwy, Crawley WA 6009, Australia
\\
$^{6}$ ARC Centre of Excellence for All-sky Astrophysics (CAASTRO), Australia
\\
$^{7}$ School of Mathematics \& Physics, University of Tasmania, Private Bag 37, Hobart, Tasmania 7001, Australia
}

\date{Accepted XXX. Received YYY; in original form ZZZ}

\pubyear{2017}

\begin{document}
\label{firstpage}
\pagerange{\pageref{firstpage}--\pageref{lastpage}}
\maketitle

\begin{abstract}
We study the mutual alignment of radio sources within two surveys, FIRST and TGSS. This is done by producing two position angle catalogues containing the preferential directions of respectively $30\,059$ and $11\,674$ extended sources distributed over more than $7\,000$ and $17\,000$ square degrees. 
The identification of the sources in the FIRST sample was performed in advance by volunteers of the Radio Galaxy Zoo project, while for the TGSS sample it is the result of an automated process presented here.
After taking into account systematic effects, marginal evidence of a local alignment on scales smaller than $2.5^\circ$ is found in the FIRST sample. The probability of this happening by chance is found to be less than $2$ per cent. Further study suggests that on scales up to $1.5^\circ$ the alignment is maximal. For one third of the sources, the Radio Galaxy Zoo volunteers identified an optical counterpart. Assuming a flat $\Lambda$CDM cosmology with $\Omega_m = 0.31, \Omega_\Lambda = 0.69$, we convert the maximum angular scale on which alignment is seen into a physical scale in the range $[19, 38]$ Mpc $h_{70}^{-1}$. This result supports recent evidence reported by Taylor and Jagannathan of radio jet alignment in the $1.4$ deg$^2$ ELAIS N1 field observed with the Giant Metrewave Radio Telescope.  The TGSS sample is found to be too sparsely populated to manifest a similar signal.
\end{abstract}


%%%%%%%%%%%%%%%%%%%%%%%%%%%%%%%%%%%%%%%%%%%%%%%%%%

%%%%%%%%%%%%%%%%% BODY OF PAPER %%%%%%%%%%%%%%%%%%

\section{Introduction}

	In the last two decades, the mutual alignment of optical linear polarizations of quasars over cosmological scales (comoving distance $\geq 100$ $h^{-1}$ Mpc) has been reported \citep{Hutsemekers1998, Lamy2001, Cabanac2005}. Since quasars are rare and non-uniformly distributed, ad hoc statistical tools have been developed over the years to study the phenomenon \citep{Jain2004, Shurtleff2013, Pelgrims2014}. Since the correlation between AGN optical polarization vectors and structural axes has been observed \citep[e.g.,][]{Lyutikov2005, Battye2009}, the coherence of the polarization vectors could be interpreted as an alignment of the nuclei themselves or alignment with respect to an underlying large-scale structure. Confirmation of this came from \cite{Hutsemekers2014}, who considered quasars known to be part of quasar groups and detected an alignment of the polarization vectors either parallel or perpendicular to the large-scale structure they belong to.
		
	Both observational results \citep[e.g.,][]{Tempel2013, Zhang2013, Hirv2016} and tidal torque analytical models \citep[e.g.,][]{Codis2015, Lee2004} suggest the alignment of galaxy spins with respect to the filaments and walls of the large-scale structure. The geometry of the cosmic web influences the spin and shape of galaxies by imparting tidal torques on collapsing proto-halos. The same mechanism might be behind both the alignment of galactic spins and polarization vectors, but the topic is still under discussion \citep{Hutsemekers2014}. The main caveats are the peculiar cosmic evolution of quasars, dominated by feedback, and the implications that an alignment on such large-scales would have for the cosmological principle \citep[see, for example,][]{Zhao2016}.
		
	\cite{Taylor2016} reported local alignment (below the $1^\circ$ scale) of radio galaxies in the ELAIS N1 field observed with the Giant Metrewave Radio Telescope (GMRT) at $610$ MHz. 
	
	Despite the known trend of radio galaxy major axes to be aligned with the optical minor axis rather than the optical
	major axis \citep[e.g.,][]{Andernach1995, Battye2009, Kaviraj2015}, the correlation between the large-scale angular momentum of the galaxy and the angular momentum axis of the material accreting towards the AGN (traced by the jets) is disputed \citep{Hopkins2012}. This makes the tidal torque interpretation of the radio jets alignment nebulous at best. On the other hand, modelling the formation of dominant cluster galaxies suggests that the spin of the black holes powering AGNs is affected by the galactic accretion history and therefore might be aligned with the surrounding large-scale structure \citep{West1994}.
	
	In this work, we attempt to corroborate and extend the results obtained in \cite{Taylor2016} by studying the alignment of radio sources in the maps of the radio sky provided by the two surveys: Faint Images of the Radio Sky at Twenty-centimeters \citep[FIRST;][]{Becker1995a} and TIFR GMRT Sky Survey\footnote{Website: http://tgssadr.strw.leidenuniv.nl/} \citep[TGSS;][]{Intema2016}. In Section~\ref{sec:SS} we construct two catalogues that contain the orientations and coordinates of resolved radio sources. In Section~\ref{sec:SA} we present the statistical instruments we make use of, based on those developed by \cite{Bietenholz1986} and \cite{Jain2004} for the study of quasar optical polarizations. In Section~\ref{sec:RES} we discuss the results of the analysis.
	
	In appendix~\ref{sec:shear} a more sophisticated approach to the study of alignment is presented. The statistics used in there do not however return any significant result.

\section{Sample Selection}
	\label{sec:SS}
	The November 2015 alpha version of the Radio Galaxy Zoo consensus catalogue lists the properties of $85\,151$ radio sources distributed primarily over the footprint of two surveys: FIRST and Australia Telescope Large Area Survey (ATLAS) \citep{Norris2006}. The classification was performed by volunteers, who were presented with radio images from these surveys and the corresponding infrared fields observed by the Wide-field Infrared Survey Explorer \citep[WISE;][]{Wright2010}. They were then asked to match disconnected components corresponding to the same source and recognize the infrared counterpart. A more detailed description of the project is available in \cite{Banfield2015}. 

	The Radio Galaxy Zoo represents a natural choice for our statistical analysis. Whereas components belonging to the same source are usually recognized through self-matching (i.e., cross-matching the source catalogue with itself to identify sources at a certain distance from each other) or human selection, we rely on the additional information provided by human inspection to increase the reliability of the results. Furthermore, the $5\arcsec$ nominal resolution of the FIRST images implies a high number of resolved sources, for which a preferential direction can be defined. Lastly,  the survey covers an area of about $10\,000$ square degrees and allows us to infer general properties of the radio sky, instead of a local statistical anomaly.

	For our second sample, based on the TGSS Alternative Data Release 1, no human-made classification is available. In its place, we opt for automated self-matching. The TGSS ADR1 is based on an independent reprocessing of an original 150 MHz GMRT survey performed between 2010 and 2012 and the corresponding source catalogue, released in 2016, covers $99.5$  per cent of the sky north of $-53^\circ$ declination. A more detailed description is available in \cite{Intema2016}.
	
	\subsection{Radio Galaxy Zoo}
	
			\begin{figure}
				\centering
				\includegraphics[width=0.45\textwidth]{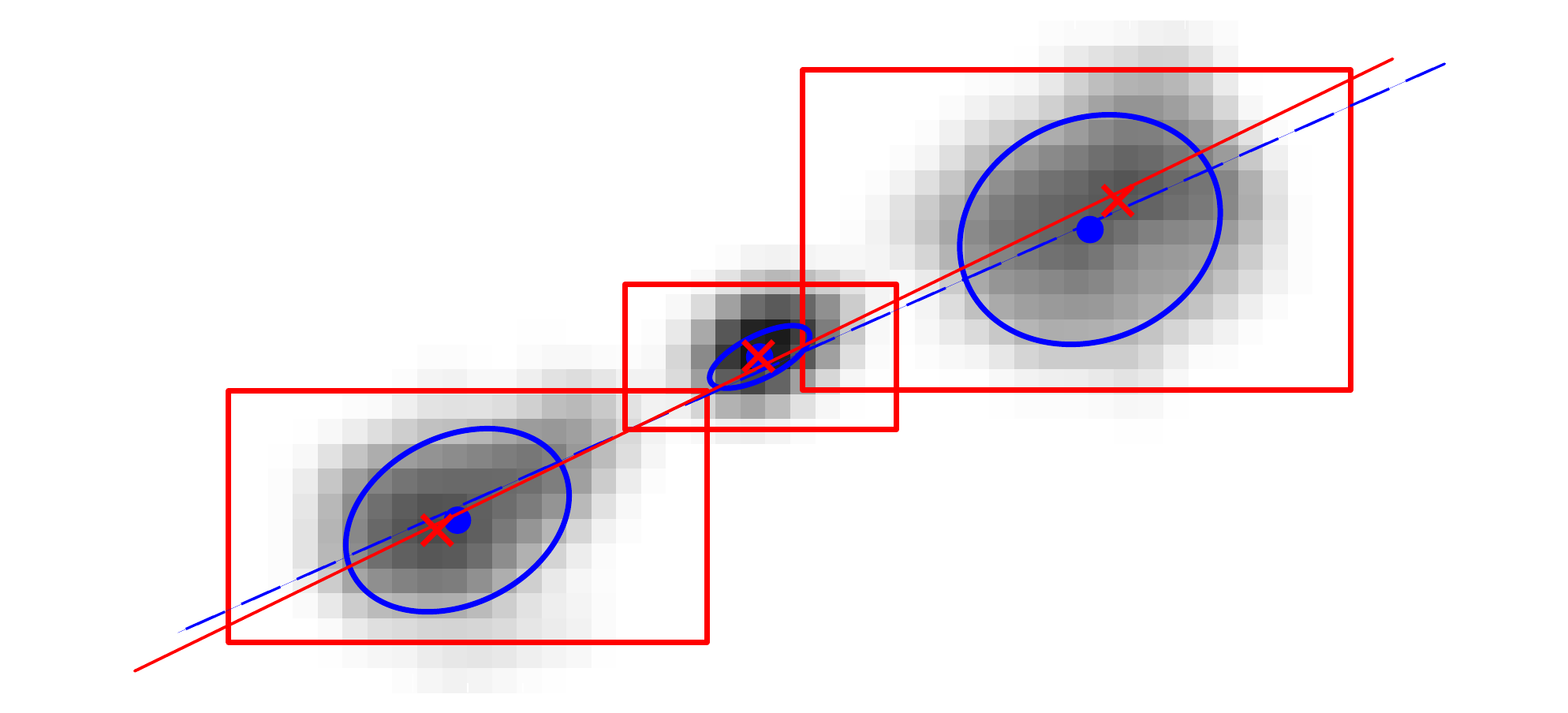}
				\caption{FIRST image for a typical source with morphological features superimposed. The angular extent of the source is about $\mathbf{1\arcmin10\arcsec}$. The red boxes identify the components provided by the Radio Galaxy Zoo, with the crosses indicating surface brightness peaks. The blue ellipses have major and minor axes equal to the FWHM of the fitted Gaussian model in the FIRST catalogue, and the dots are their centres. The red and dashed blue lines are the results of the orthogonal distance regression for the dots and the crosses respectively. In this particular case two or more surface brightness peaks are present and the position angle is extracted from the slope of the red line (see text for more details).}
				\label{fig:exall}
			\end{figure}	
			
			We select extended sources from the Radio Galaxy Zoo consensus catalogue and extract an elongation direction for each of them.We describe this direction with a position angle, defined as the angle east of north in the range $[-\pi/2, +\pi/2]$ between the direction itself and the local meridian. 
			
			To perform the selection and constrain the orientation, we rely on the quantities contained in the November 2015 alpha version of the Radio Galaxy Zoo consensus catalog and, occasionally, on the official FIRST catalogue presented in \cite{Helfand2015b}, version \textsc{14dec17}. Figure~\ref{fig:exall} presents these quantities in graphic form. From the Radio Galaxy Zoo catalogue we extract the areas covered by components belonging to the same source and the peak positions of the source surface brightness contained in these regions (peaks hereafter). From the FIRST catalogue we extract the Gaussian model of the source brightness contained inside the same areas. 
			An additional quantity provided by the Radio Galaxy Zoo for every morphological classification is the consensus level. This is defined as the fraction of users who voted for the specific components configuration and, in this analysis, it is used to rank distinct classifications of the same object. 
			
			Depending on the available data, different procedures are employed to extract the position angle. We define three sub-samples:
			
			\begin{description}
			\item a)  If two or more surface brightness peaks are present for a given source, we define the position angle as the slope of the orthogonal distance linear regression of the peaks, weighted according to their flux densities. Around $80$ per cent of the selected sources belong to this category. An example of such a source is provided in Figure~\ref{fig:exall}.

			\item b) For sources with only one surface brightness peak in the Radio Galaxy Zoo catalogue, but multiple Gaussian models in the FIRST catalogue, we rely completely on the latter.
			This occurs when components are not seen as separated in the Radio Galaxy Zoo because of the particular automated choice of contour levels. For this sub-sample the centres of the FIRST ellipses, weighted by their integrated flux, are fitted.
			
			\item c) 	If only one surface brightness peak is detected and the FIRST catalogue recognizes only one source inside the single component radio galaxy, we rely on the Gaussian model of the FIRST catalogue and we define the direction as the position angle of the fitted ellipse. In this case, the source must comply with a total of four criteria.
			
			First, sources must meet the conditions required to be included in the Radio Galaxy Zoo sample and be presented to the volunteers. These are aimed at selecting resolved sources with a high signal-to-noise ratio:
			
			\begin{equation}
			\frac{S_{\rm{peak}}}{S_{\rm{int}}} < 1.0 - \left( \frac{0.1}{\log S_{\rm{peak}}}\right) \textrm{ and } SNR>10,
			\label{eq:cond1}
			\end{equation} 
			where $S_{\rm{peak}}$ is the peak brightness in mJy beam$^{-1}$, $S_{\rm{int}}$ is the integrated flux density of the source in mJy and $SNR$ is the signal-to-noise ratio \citep{Banfield2015}.
			
			Secondly, we introduce two additional criteria. The minor axis $m$ of the fitted elliptical Gaussian model should be larger than $2\arcsec$ and the deviation of the ratio between the major and minor axis $r$ from unity should be highly significant:

			\begin{equation}
			m > 2\arcsec
			\textrm{ and }
			r > 1 + 7\sigma_r.
			\label{eq:cond2}
			\end{equation}
			
			The error on the major and minor axis ratio is overestimated by the quadratic sum

			\begin{equation}
			\sigma_r = r\sqrt{\left(
				\frac{\sigma_m}{m}
				\right)^2
				+
				\left( 
				\frac{\sigma_m}{M}
				\right)^2},
			\end{equation}

			where $\sigma_m$ is the empirical uncertainty on both the fitted minor axis $m$ and major axis $M$. The four conditions, \eqref{eq:cond1} and \eqref{eq:cond2}, select extended sources for which an elongation is clearly recognizable.
			
			\end{description}
	
			When both multiple Gaussian models and multiple flux density peaks are available, we choose to prioritize the peaks over the centres. Figure~\ref{fig:exall} provides an example of how the difference between the two fitted position angles is usually small.

			\begin{figure}
				\centering
				\includegraphics[width=0.5\textwidth]{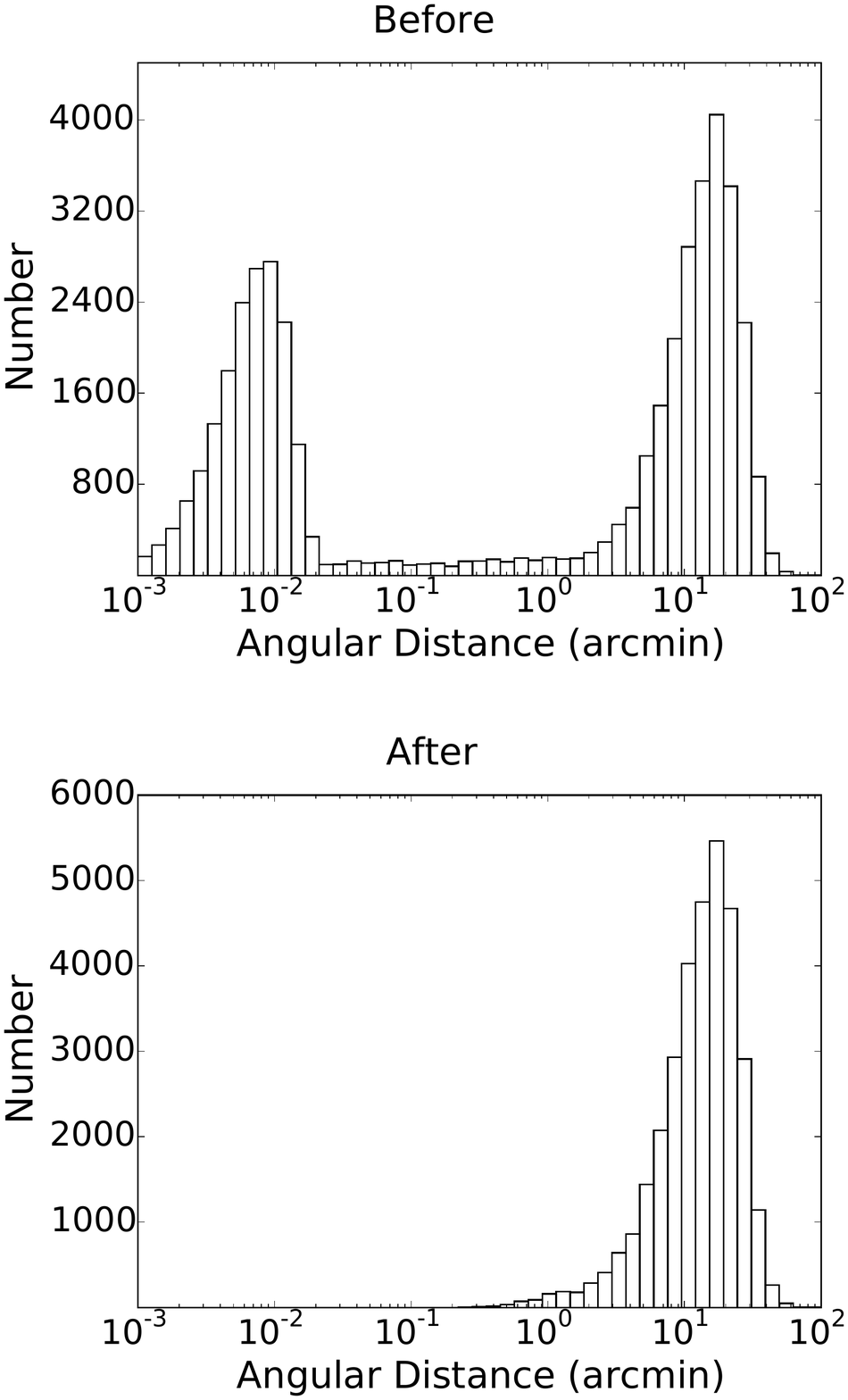}
				\caption{Distribution of the angular distance between a source in the Radio Galaxy Zoo sample and its closest neighbour, before and after filtering duplicates.}
				\label{fig:mindist}
			\end{figure} 
			
			The release of the Radio Galaxy Zoo consensus catalogue used here includes every classification performed by the volunteers. Because of this, a single source might appear multiple times with different classifications. 
            To filter these duplicate entries we focus our attention on all the recognized components. For every set of overlapping components, we filter out all of the sources they belong to, except for the one with highest consensus level. The effect of this selection process can be seen in Figure~\ref{fig:mindist}, where we plot the distribution of the distances between every source and its closest neighbour. While a natural amount of clustering is expected, we find that almost half of the sources have an extremely close neighbour -- a probable duplicate. After we apply our filter the peak around $0\farcs6$ disappears.

\begin{figure}
		 		\centering
		 		\includegraphics[width=0.5\textwidth]{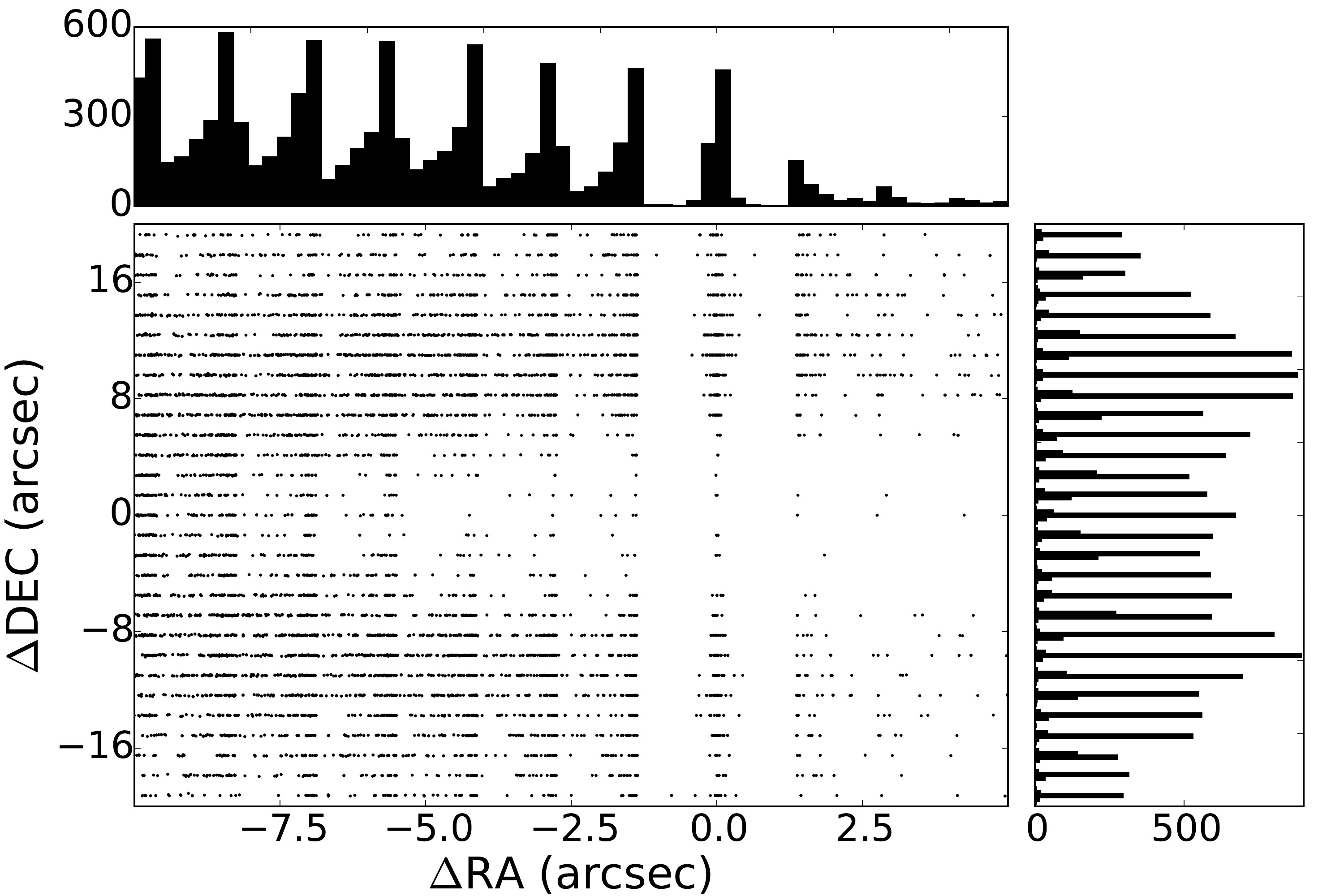}
		 		\caption{Relative peak positions for entries classified as containing two peaks. Discretization is evident in the collapsed distributions.}
		 		
		 		\label{fig:PP}
		 	\end{figure}

            A second systematic effect inherited from the Radio Galaxy Zoo is the quantisation of the peak positions. To clearly discern its importance, we limit our attention to the sources classified as containing only two peaks and we plot the differential right ascension and declination of every pair (Figure~\ref{fig:PP}). Discretisation is more noticeable in the vertical axis, but a $1\farcs4$ binning effect is visible in both directions. The presence of pixels is caused by numerical approximations in the implementation of the World Coordinates System (WCS).  
			In our analysis, this grid-like disposition of the peaks implies discrete values of the associated position angles. To obtain a continuous distribution of the angles, we smooth out the peak positions by adding a uniformly random value in the range $[-0\farcs7, +0\farcs7]$ to both coordinates before performing the linear regression. This process pushes the influence of the effect to sub-pixel scales, eliminating its impact on the present study. However, further investigation is needed to constrain its causes. 

\begin{figure}
				\centering
				\includegraphics[width=0.45\textwidth]{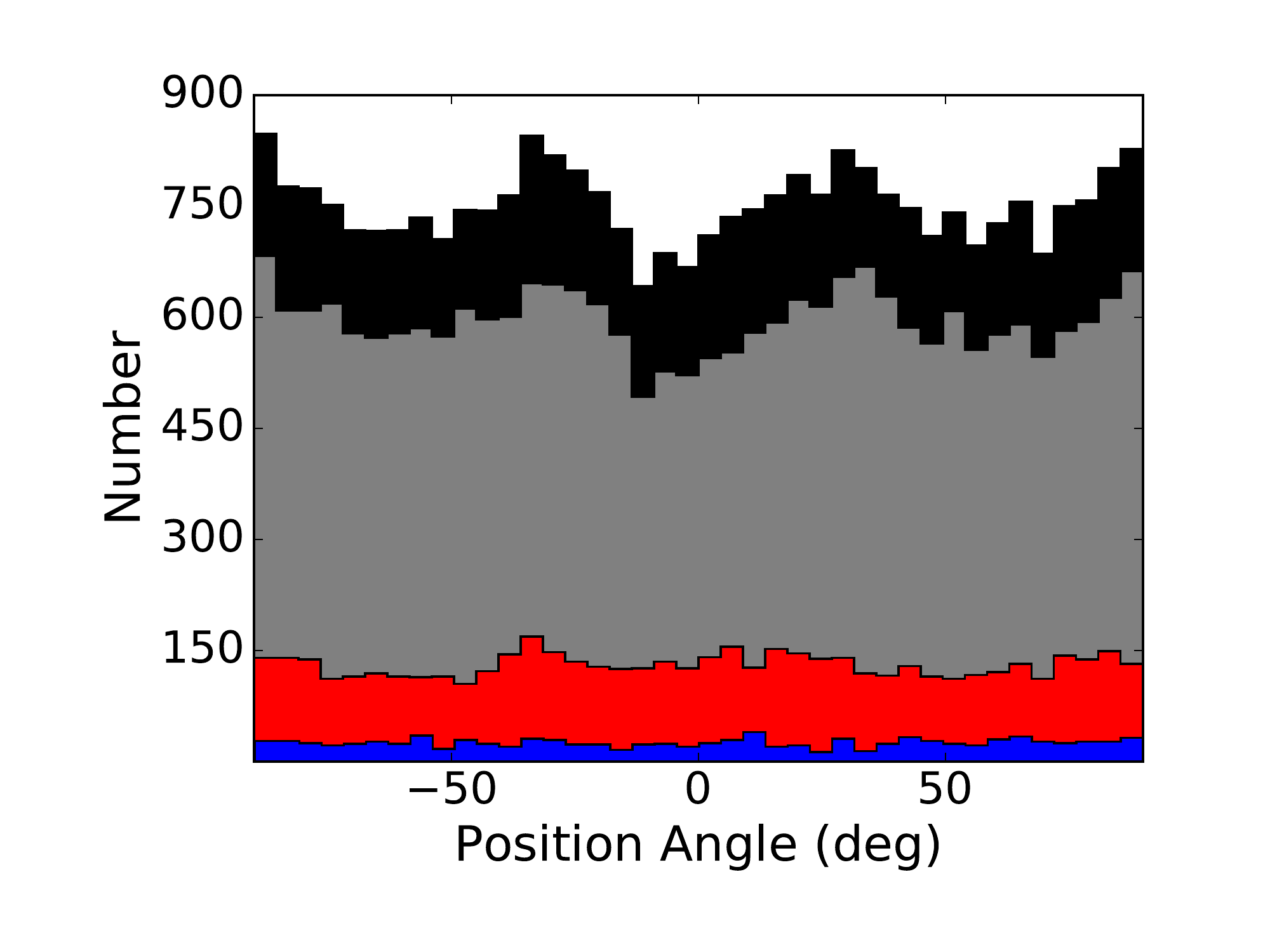}
				\caption{Position angle distribution of the Radio Galaxy Zoo selection. On top of the total distribution (topmost histogram) the plot contains the distributions of the three sub-samples. From top to bottom: (a) in grey, (b) in red, and (c) in blue. A trimodal systematic effect is visible in the first two.}
				
				\label{fig:RGZPAdist}
			\end{figure}

            For the sake of consistency, the final sample presented in Figure~\ref{fig:RGZPAdist} excludes ATLAS sources and it is limited only to FIRST sources. For the same reason, we also exclude every source positioned above RA $20$ hr and below $4$ hr, since half of the observations in this region were performed after the observing array transitioned to the new JVLA configuration \citep{Helfand2015b}. 
            
			Finally, notice how the original Radio Galaxy Zoo selection in Eq.~\eqref{eq:cond1} does not include an explicit cut for artefacts. During the first run of the Radio Galaxy Zoo classification, the volunteers were presented with $3\arcmin \times3\arcmin$ fields. This corresponds to a maximum distance of $ 3\arcmin \sqrt{2} \approx 4\arcmin 12\arcsec$ between two components. To quantify the contamination from artefacts in our sample, we make use of the column $P(S)$ of the official FIRST catalogue, which indicates the probability of a source to be a sidelobe. We cross-matched our selection with the FIRST catalogue using a search radius of $4\arcmin 12\arcsec$ and we verified that 134 selected sources are part of a field containing possible sidelobes satisfying the condition $P(S)>0.1$. In principle, these artefacts might be recognized as components and influence the value of the position angles. Because of this, we exclude sources with $P(S)>0.1$ from our final Radio Galaxy Zoo sample.
            
            In Figure~\ref{fig:RGZPAdist} we plot the final distribution of the extracted position angles, together with the distributions for the three classes of sources. While we would expect these to be uniform, three peaks are visible around $30^\circ$, $-30^\circ$ and $90^\circ$. In these three directions we recognize the typical pattern that results from the three arms of the observing radio interferometer --- the Very Large Array (VLA). The same effect is visible in the FIRST images and is discussed in \cite{Helfand2015b}, where a three-directional pattern is present in the distribution of the sidelobes around bright sources. The existence of preferential angles may be related to the brightness of the weaker components, although a more detailed analysis would be required to quantify this effect. This will not affect our analysis as long as the effects are non-local.

            A similar pattern is discussed also in other analyses \citep[e.g.,][]{Chang2004, White2007, Demetroullas2015} based on the FIRST survey, where the effect is recognized as non position-dependent. Snapshot surveys are commonly affected by an anisotropic point spread function (PSF) and the connection to the interferometer geometry suggests this origin.
            \cite{Helfand2015b} underlines that particular care was taken in ensuring a constant PSF throughout the different observation epochs of FIRST. In particular, since the hour angle of observation affects the orientation of the pattern in the cleaned images, $90\%$ of the observations were acquired within $1.4$ hr of the local meridian.	
			
			The non-locality of the effect is verified by partitioning the data by both right ascension and declination in four equally populated quadrants. Pairwise, the four position angle distribution are found to be consistent with each other using two-sample Kolmogorov-Smirnov tests. 
            
            This first position angle catalogue contains $30\,059$ sources distributed over an area of about $7\,000$ square degrees, resulting in a number density $\sim4$ deg$^{-2}$.

		\subsection{TGSS Alternative Data Release}
		
\begin{figure}
				\centering
				\includegraphics[width=0.5\textwidth]{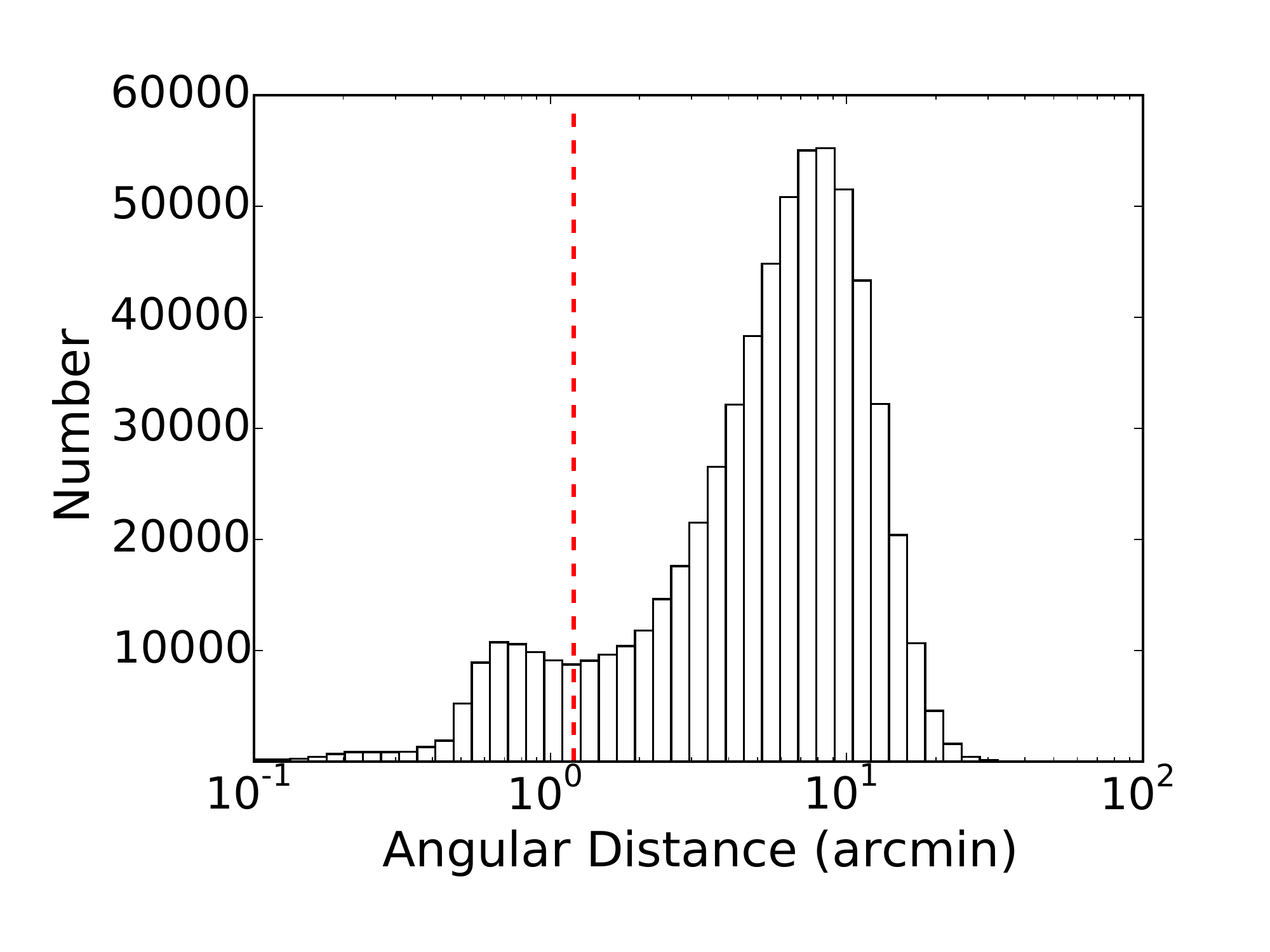}
				\caption{Distribution of the angular distance between a source in the TGSS catalogue and its closest neighbour. The dashed red line marks the value $1\arcmin12\arcsec$.}
				
				\label{fig:mindistTGSS}
			\end{figure}

			As opposed to the Radio Galaxy Zoo sample, this second position angle sample is based on the product of an automated source extractor. The nominal resolution of $25\arcsec$ for the TGSS images implies a lower number of extended sources with significant elongation compared to FIRST. However, the relatively steep spectrum of radio galaxy lobes and the sensitivity to extended sources of the GMRT allow TGSS to trace the lobes better than FIRST.  Hence, we focus our attention on the identification of double-lobed sources. 
			In Figure~\ref{fig:mindistTGSS} we plot the distance between each entry in the TGSS catalogue and its closest neighbour. The rightmost peak is due to the distribution of uncorrelated radio sources, while the lower peak on the left is caused by multi-component sources. The plot suggests an average distance of $1\arcmin$ between the components of a source of the latter type. A peak around the angular scale of $1\arcmin$ is not present in the Radio Galaxy Zoo catalogue because the pairing was already performed by the volunteers during the classification process.

			We select radio galaxy candidates by self-matching the catalogue with a search radius $1\arcmin12\arcsec$ and imposing a maximum ratio of $10$ between the total fluxes of the two components \citep{VanVelzen2014}.
			To be part of the final sample both components of the pair need to satisfy additional constraints: (1) isolated (i.e., matched only to each other) (2) $SNR>10$. The position angle is then simply that of the line connecting the two components. The search radius we chose corresponds to the local minimum marked in Fig.~\ref{fig:mindistTGSS}. A larger value would introduce an artificial contamination in our double-lobed source catalogue, while a lower value would mean losing part of the genuine sources.
			
			We decide to limit our sample to a portion of the northern hemisphere to minimize the effects of an anisotropic PSF. \cite{Intema2016} reports the synthesized beam to be circular for pointings at declination higher than the GMRT latitude --- about $19^\circ$. Even between declinations of $10^\circ$ and $19^\circ$ the beam is still circular to within $1\%$. Therefore, our final TGSS sample includes only sources with declination above $10^\circ$.
			
\begin{figure}
				\centering
				\includegraphics[width=0.45\textwidth]{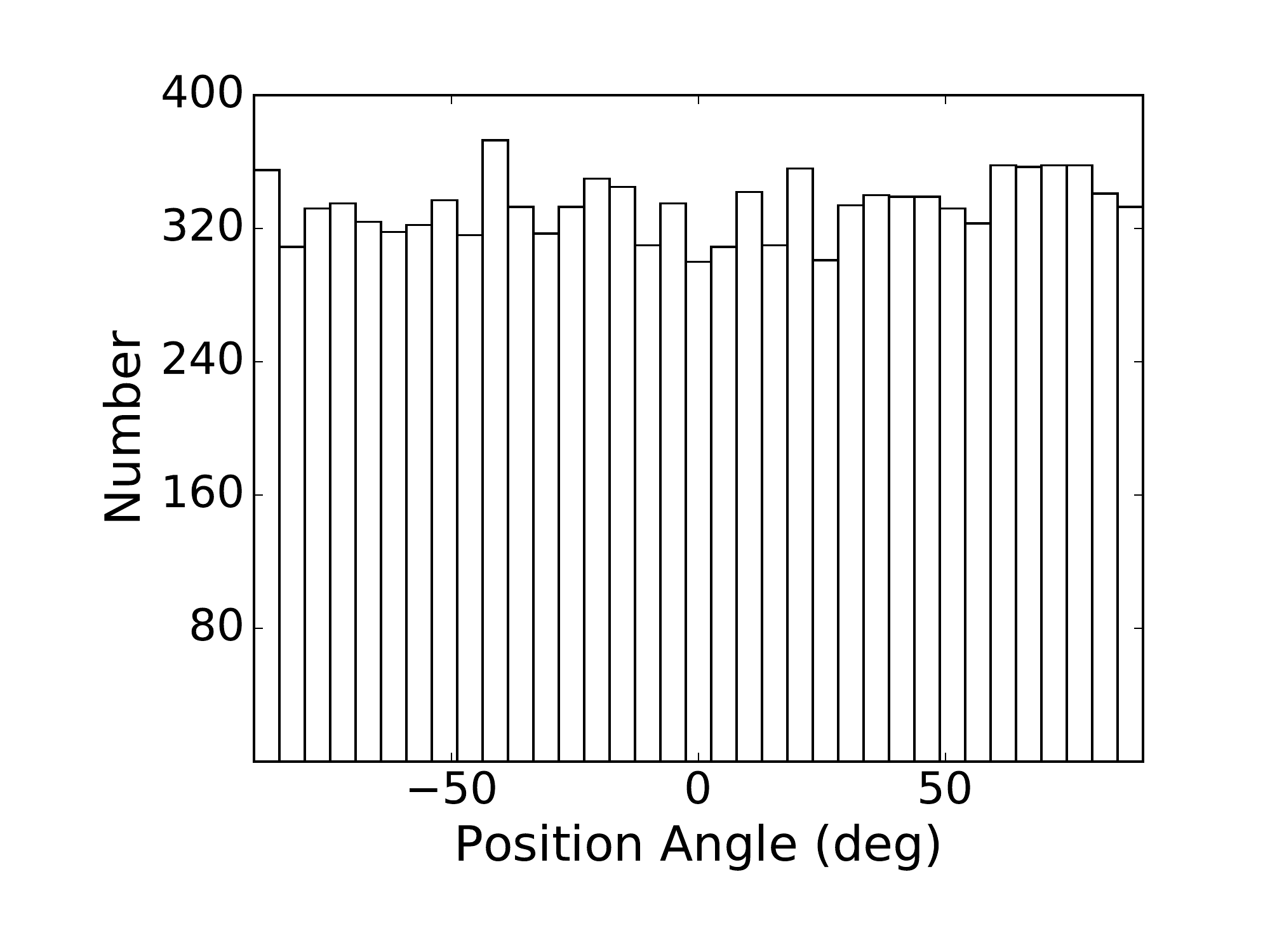}
				\caption{Position angle distribution of the TGSS selection. Obvious systematic effects are not present.}
				
				\label{fig:TGSSPAdist}
			\end{figure}

			Figure~\ref{fig:TGSSPAdist} shows the position angle distribution of the final TGSS sample. This second position angle catalogue contains $11\,674$ sources distributed over an area of about $17\,000$ square degrees, resulting in a number density $\sim0.7$ deg$^{-2}$. We notice that unlike for the FIRST survey, no particular care was taken with respect to the PSF and its consistency throughout different pointings. However, the  complex geometry of the interferometer and longer integration times compared to FIRST result in a PSF less prone to systematic effects. 
			Table~\ref{tab:surveys} compares the different surveys and samples featured in this section. The difference between the number of sources in the two catalogues produced in this section is due to the different nature of the original surveys and the source selection process. While $85\%$ of the sources in the RGZ sample have size larger than the TGSS resolution ($25\arcsec$), only $55\%$ of them are larger this threshold and have exactly two surface brightness peaks.
			
			We can use the RGZ catalogue to predict the size of the TGSS one. If we account for the different frequencies ($1.4$ GHz for FIRST and $150$ MHz for TGSS) by adopting a nominal spectral index equal to $0.9$ \citep{Vollmer2010} and keeping in mind the sky coverage and angular resolution differences, we find that about $10^4$ sources are expected to be selected by our algorithm. This number is in line with the $11\,674$ sources found in our selection.
			
\begin{table*}
\caption{Comparison between the different samples and source catalogs discussed in this paper.}
\label{tab:surveys}
\begin{tabular}{lccccccc}
\hline
Name & Frequency & Median RMS & SNR & Number of & Minimum & Sky & Median Redshift \\
     &           & Noise      & Threshold  & Sources   &  Resolution & Fraction & $68\%$ interval \\
     &           &[mJy beam$^{-1}$] &       &           &            &          &                  \\ 

\hline
				FIRST\,$^a$ &
				$1.4$ GHz & 
				$0.15$& 
				$5$&
				$946\,432$& 
				$5\arcsec\times5\arcsec$& 
				$26\%$&
				$2.2\pm0.9$\,$^b$ 
				\\ 
				Radio Galaxy Zoo\,$^c$  & 
				$1.4$ GHz & 
				$0.15$ & 
				$10$& 
				$82\,187$&  
				$5\arcsec\times5\arcsec$& 
				$22\%$& 
				$0.47_{-0.15}^{+0.21}$\,$^d$ 
				\\
				Radio Galaxy Zoo processed\,$^e$ & 
				$1.4$ GHz &  
				$0.15$& 
				$10$& 
				$30\,059$  &
				$5\arcsec\times5\arcsec$& 
				$19\%$& 
				$0.47_{-0.15}^{+0.20}$\,$^d$ 
				\\
				TGSS\,$^f$& 
				$150$ MHz & 
				$3.5$& 
				$7$& 
				$623\,604$ & 
				$25\arcsec\times25\arcsec$& 
				$90\%$& 
				$-$
				\\
				TGSS processed\,$^e$ & 
				$150$ MHz & 
				$3.5$& 
				$10$& 
				$11\,674$ & 
				$25\arcsec\times25\arcsec$& 
				$42\%$& 
				$-$ \\
\hline
\multicolumn{8}{l}{ $^a$ \cite{Helfand2015b}} \\
\multicolumn{8}{l}{ $^b$ Mean redshift with $68\%$ confidence levels from \cite{Chang2004}}	\\
\multicolumn{8}{l}{ $^c$ \cite{Banfield2015}} \\
\multicolumn{8}{l}{ $^d$ Only $30\%$ of the sample has a human-matched optical counterpart with known redshift} \\
\multicolumn{8}{l}{ $^e$ The selection process, aimed at selecting resolved sources to use in this study, is detailed in Section~\ref{sec:SS}} \\
\multicolumn{8}{l}{ $^f$ \cite{Intema2016}} \\
\end{tabular}
\end{table*}

\section{Statistical Analysis}
	\label{sec:SA}
	\subsection{Parallel Transport}	
		\label{sec:PT}

		The position angle is a directional quantity defined in the point of the celestial sphere where the corresponding source lies. In order to perform the calculation of the misalignment angle between two directions on a sphere, the notion of parallel transport should be introduced \citep{Jain2004}.
		
		We parametrize the sphere using spherical coordinates $(r, \theta, \phi)$ and we define in every point a natural orthonormal basis dictated by our coordinate system. This set of unit vectors is $(\mathbfit{e}_r, \mathbfit{e}_\theta, \mathbfit{e}_\phi)$, where the three elements point respectively towards the centre of the sphere, northward and eastward.	
		
		A source with position angle $\alpha$, determined up to a rotation of $\pi$ radians, can be identified with the unit vector
		
		\begin{equation}
			\mathbfit{v} = \cos\alpha \; \mathbfit{e}_\theta + \sin \alpha \; \mathbfit{e}_\phi
			\label{eq:v}
		\end{equation}
		
		Since the projection along the line of sight is unknown, we fix this vector to be tangent to the sphere at the point of definition.  
		The vector $\mathbfit{v}$ represents a physical quantity, whereas the definition of position angle $\alpha$ depends on the choice of coordinate system. For example, if parallels and meridians were redefined with respect to a different north pole, the vectors $\mathbfit{e}_\theta$, $\mathbfit{e}_\phi$ and the position angle $\alpha$ would change. However, the vector $\mathbfit{v}$ in Eq. \eqref{eq:v} would still describe the same direction in space.
		On a sphere, parallel transport allows us to define a coordinate-invariant inner product between two vectors, by translating one of them along arcs of great circles connecting the two.  
		
		Let us consider two tangent vectors $\mathbfit{v}_1$ and $\mathbfit{v}_2$ with position angles $\alpha_1$ and $\alpha_2$, defined respectively in $P_1 = (r_1, \theta_1, \phi_1)$ and $P_2 = (r_2, \theta_2, \phi_2)$. Both of these points belong to the same unit sphere ($r_1 = r_2 = 1$). The great circle passing through them lies on a plane perpendicular to $\mathbfit{e}_s$
		
		\begin{equation}
			\mathbfit{e}_s = \frac{\mathbfit{e}_{r_1}\times \mathbfit{e}_{r_2}}{\vert \mathbfit{e}_{r_1} \times \mathbfit{e}_{r_2}\vert}
		\end{equation}
		
		We define $\mathbfit{e}_{t_1}$ and $\mathbfit{e}_{t_2}$ as the tangent vectors of this great circle in the points $P_1$ and $P_2$.
		
		\begin{gather}
			\mathbfit{e}_{t_1} = \mathbfit{e}_{s} \times \mathbfit{e}_{r_1}
			\\
			\mathbfit{e}_{t_2} = \mathbfit{e}_{s} \times \mathbfit{e}_{r_2}
		\end{gather}
		
		We call $\zeta_1$ the angle between $\mathbfit{e}_{t_1}$ and $\mathbfit{e}_{\theta_1}$. Similarly, we define $\zeta_2$ as the angle between $\mathbfit{e}_{t_2}$ and $\mathbfit{e}_{\theta_2}$. Translating the vector $\mathbfit{v}_{1}$ along the great circle maintains the angle with the local tangent vector constant and at the point $P_2$ it results in the translated vector $\mathbfit{v}_1^\prime$ with position angle
		
		\begin{equation}
			\alpha_1^\prime =  \alpha_1 + \zeta_2 - \zeta_1
			\label{eq:alphaprime}
		\end{equation}
		
\begin{figure}
	\centering
	\includegraphics[width=0.45\textwidth]{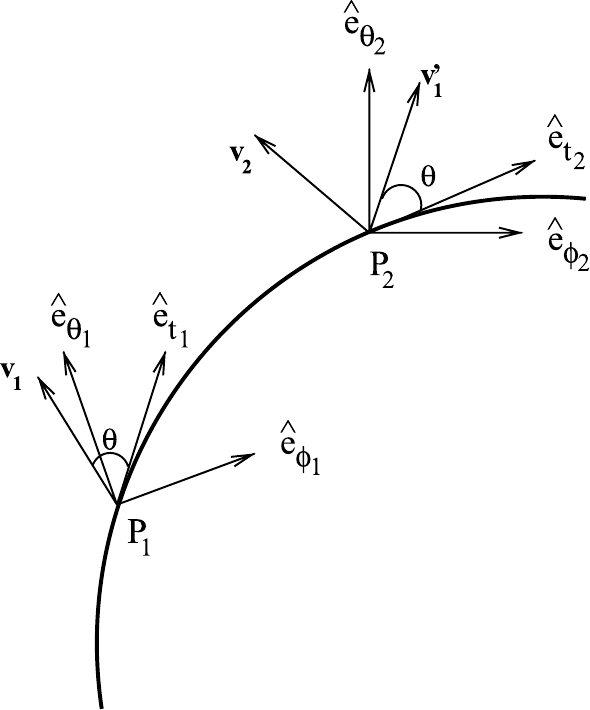}
	\caption{Two dimensional schematic illustration of parallel transport. The figure displays the arc of great circle passing through the points $P_1$ and $P_2$, with $\mathbfit{e}_{t_1}$ and $\mathbfit{e}_{t_2}$ tangent vectors to curve in these points. Notice that the angle $\theta$ between the tangent vector and $\mathbfit{v}_{1}$ is kept constant when $\mathbfit{v}_{1}$, located at $P_1$, is translated along the curve to the point $P_2$. The figure is taken from \citet{Jain2004}, their figure 1, with the author's permission.}
	\label{fig:PT}
\end{figure}

		Figure~\ref{fig:PT} depicts the vectors involved in the operation. With this in mind, we define the generalized dot product between $\mathbfit{v}_{1}$ and $\mathbfit{v}_{2}$ as the following
		
		\begin{equation}
			\mathbfit{v}_{1} \odot \mathbfit{v}_{2} = \vert \mathbfit{v}_{1} \vert
			\vert \mathbfit{v}_{2} \vert \cos (\alpha_1 - \alpha_2 + \zeta_2 - \zeta_1)
		\end{equation}
		
		Since our dataset is purely directional, we have $\vert \mathbfit{v}_{1} \vert = \vert \mathbfit{v}_{2} \vert = 1$. For the same reason, the inner product is written using the following simplified notation
		
		\begin{equation}
		(\alpha_1, \alpha_2) =  \cos [2(\alpha_1 - \alpha_2 + \zeta_2 - \zeta_1)]
		\label{eq:innerproduct}
		\end{equation}
		
		The factor two is introduced so that the argument of the cosine ranges over the full $-\pi$ to $+\pi$, \citep{Bietenholz1986}. By definition $(\alpha_1, \alpha_2) \in [-1, 1]$, where $+1$ indicates perfect alignment \citep{Jain2004} and $-1$ implies perpendicular directions.

	\subsection{Angular Dispersion}
		\label{sec:S}
		Given the $i-$th source, we consider the $n$ sources closest to it (including itself). We call $d_{i,n}$ the dispersion function of their position angles.
		
		\begin{equation}
			d_{i, n}(\alpha) = \frac{1}{n}\sum_{k=1}^{n} (\alpha, \alpha_k)
			\label{eq:d}
		\end{equation}
		
		This quantity is a function of a position angle $\alpha$ located at the point where the $i-$th source lies. We call $\alpha_{\rm{max}}$ the position angle that maximizes the dispersion, which assumes the value
		
		\begin{equation}
			d_{i, n}\big|_{\rm{max}} =  \frac{1}{n} \left[ 
			\left( \sum_{k=1}^{n} \cos 2\alpha_k^\prime\right)^2
			+
			\left( \sum_{k=1}^{n} \sin 2\alpha_k^\prime\right)^2
			\right]^{1/2},
		\end{equation}
		
		where $\alpha_k^\prime$ was defined in Eq.~\eqref{eq:alphaprime} and corresponds to the value of the original position angle $\alpha_k$ after being transported in the $i-$th position. Following \cite{Jain2004}, we regard this maximal value as the measure of the dispersion of the $n$ sources and $\alpha_{\rm{max}}$ as their mean direction. The maximum value allowed for the dispersion is $d_{i, n}|_{\rm{max}} = 1$, corresponding to perfect alignment of the sources. The coordinate-invariance of the inner product (Eq. \ref{eq:innerproduct}) extends to the dispersion.

		For a sample of $N$ sources we fix a number of nearest neighbours $n$ and we derive the set of dispersions.
		
		\begin{align}
			\{d_{i, n}\big|_{\rm{max}}\} && i =1, \dots, N
		\end{align}
		
		For this set we define the following statistics
		
		\begin{align}
			S_{n} = \frac{1}{N} \sum_{i=1}^{N} d_{i, n}\big|_{\rm{max}},
			\label{eq:S}
		\end{align}

		corresponding to the mean dispersion. $S_n$ measures the average position angle dispersion of the sets containing every source and its $n$ neighbours.  
		If the condition $N \gg n \gg 1$ is satisfied, then $S_{n}$ is expected to be normally distributed. \citeauthor{Jain2004} reports the following form for its variance
		
		\begin{equation}
		\sigma_n^2 = \frac{0.33}{N},
		\label{eq:sigmaest}
		\end{equation} 
		
		where $N$ is the total number of sources in the sample. 
		The quantity $S_n$ can be employed for different values of $n$, although these different measurements are not independent. Because the dispersion $d_{i, n}$ is defined in Eq. \eqref{eq:d} as an average of the $n$ closest neighbours, the presence of a positive alignment for $n^\ast$ neighbours implies a preferential positive signal for every $n>n^\ast$.
		
		The deviation of the dispersion $d_{i, n}|_{\rm{max}}$ from its mean value is not normalized, but is found to be $\propto 1/\sqrt{n}$ \citep{Jain2004}. This is mirrored by $S_n$
		
		\begin{equation}
			S_n \propto \frac{1}{\sqrt{n}}
			\label{eq:Sest}
		\end{equation}
		
		To remove this spurious dependence, we will write the measurements of $S_n$ as one-tailed significance levels when considering multiple values of $n$
	
		\begin{equation}
			S.L. = 1-\Phi \left( \frac{S_n - \braket{S_n}_{MC}}{\sigma_n}\right),
			\label{eq:SL}
		\end{equation}
		
		where $\Phi$ is the cumulative normal distribution function and $\braket{S_n}_{MC}$ is the expected value for $S_n$ in absence of alignment, found through Monte Carlo simulations.
		We then employ the following approximate scale: $\log$ S.L. $< -3.5$, very strong alignment;  $-2.5>\log$ S.L. $> -3.5$, strong alignment; $-1.5>\log$ S.L. $> -2.5$ weak alignment.
		
		For every source (labelled by $i$) we define $\varphi_{i, n}$ as angular radius of the circle containing its $n$ neighbours. We can then define the following set:
		
		\begin{align}
		\{\varphi_{i, n}\} && i =1, \dots, N
		\label{eq:phii}
		\end{align}
		
		The distribution of this set provides information about what angular scale a particular $S_n$ probes. For our purposes we will refer to its median $\tilde{\varphi}(n)$ and the $68\%$ interval around it.

		\subsection{Random Datasets}
		\label{sec:RandomDatasets}
		
		To estimate the uncertainties and the significance of a given measurement we use simulated data sets containing only noise. The random data sets ($1\,000$ in total) are generated by shuffling the position angles among different sources to ensure that every configuration is affected by the same position angle distribution and survey geometry. 
		
		For a binned or sampled quantity $W_k$ $k\in \{1\dots N_{bins}\}$ we estimate the covariance matrix as
		
		\begin{equation}
		\Sigma^2_{ij} = \braket{(W_i - \braket{W_i}_{MC} )\cdot (W_j - \braket{W_j}_{MC})}_{MC},
		\label{eq:cov}
		\end{equation}
		
		where all the averages are computed over multiple simulations.
		
		For a multivariate Gaussian random vectors $\bm{x}$ with expected mean $\bm{\mu}$ and covariance matrix $C$ of rank $k$, the $\chi^2$ test is generalized using the Mahalanobis distance squared
		
		\begin{equation}
		d^2 = (\bm{x} - \bm{\mu})^T C^{-1} (\bm{x} - \bm{\mu}),
		\end{equation}
		
		which is chi-square distributed with $k$ degrees of freedom. In our analysis, we define the components of vector $\bm{W}$ as the measurements of the statistics $W$ performed on different scales. We then use as Mahalanobis statistics the following expression:
		
		\begin{equation}
		d^2 = (\bm{W}-<\bm{W}>_{MC})^T (\Sigma^2)^{-1} (\bm{W}-<\bm{W}>_{MC})
		\label{eq:Maha}
		\end{equation}
		
		The alignment analyses performed by \cite{Jain2004, Hutsemekers2014, Taylor2016} are based on statistical tests similar to the position angle/polarization vector mean dispersion $S_n$ defined in Eq. \eqref{eq:S}.  None of the above references take covariance into account when estimating the significance level of the measured dispersion as a function of the angular scale. In this study, the Mahalanobis statistics measures deviation from the noise by taking covariance into account.
	
\section{Results}
	\label{sec:RES}
 
	Unless stated otherwise, in this section we assume as our null hypothesis the absence of spatial coherence in the orientations of radio sources. 
 
\begin{figure}
			\centering
			\includegraphics[width=.45\textwidth]{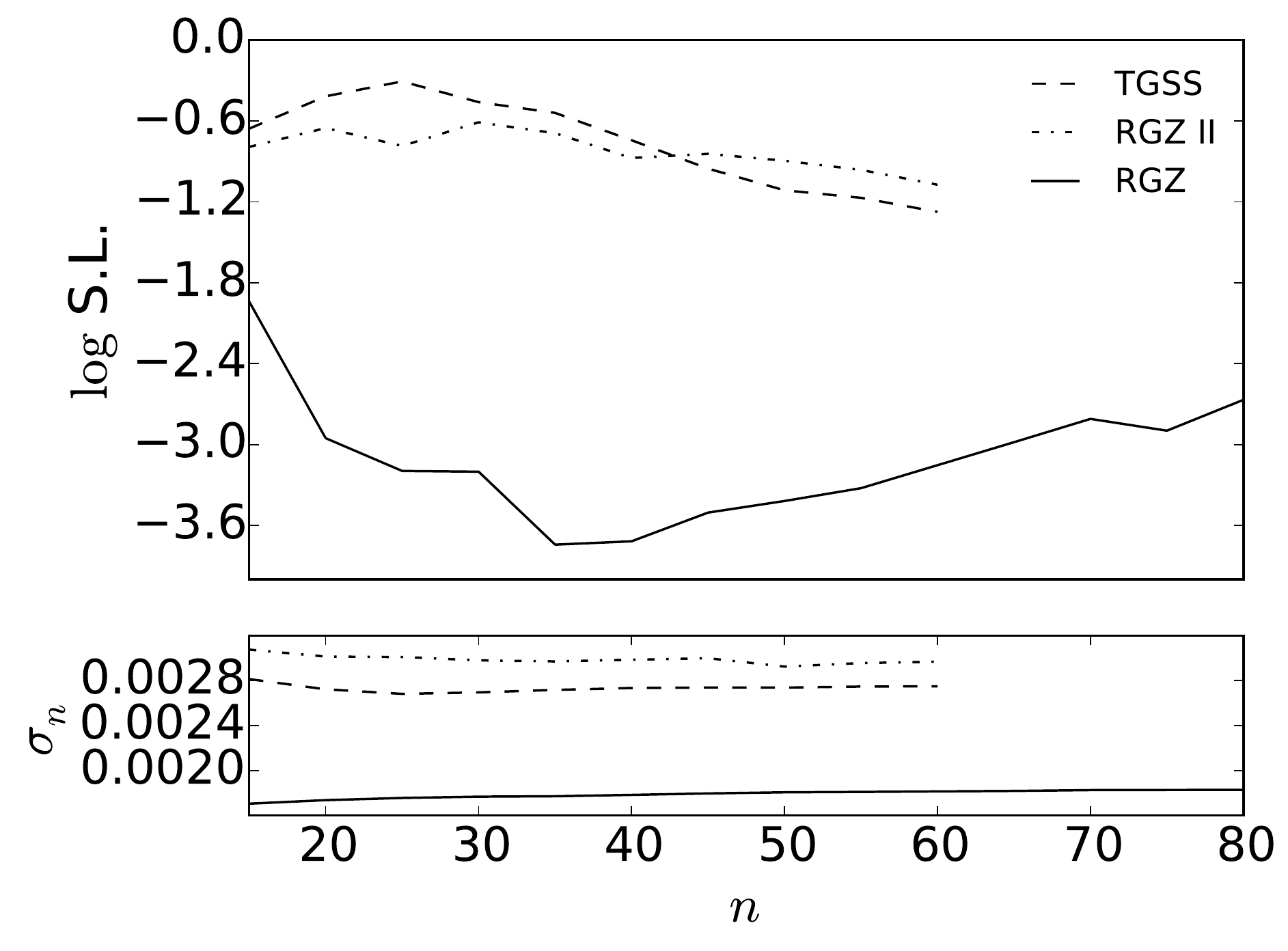}
			\caption{Logarithm of the significance level (S.L.) of the statistics $S_n$ as a function of the number of neighbours $n$ applied to three samples (see text for details). The sample standard deviation of the simulated datasets is also plotted.}
			\label{fig:Sn}
		\end{figure}

	In Fig.~\ref{fig:Sn} we plot the significance levels (S.L.) of the angular dispersion statistics $S_n$ for three different position angle samples: (1) Radio Galaxy Zoo or RGZ (2) TGSS (3) A subset of the Radio Galaxy Zoo sample, or RGZ II. This last one is designed to mimic the source count and number density of the TGSS sample, by randomly eliminating two thirds of the sources in the RGZ sample. This results in a reduced number count of $10\,088$ and a number density of about $1.5$ deg$^{-2}$. We use this dataset to also confirm that the relations \eqref{eq:Sest} and \eqref{eq:sigmaest} are confirmed up to a margin of $10\%$.
	
\begin{figure}
			\centering
			\includegraphics[width=.45\textwidth]{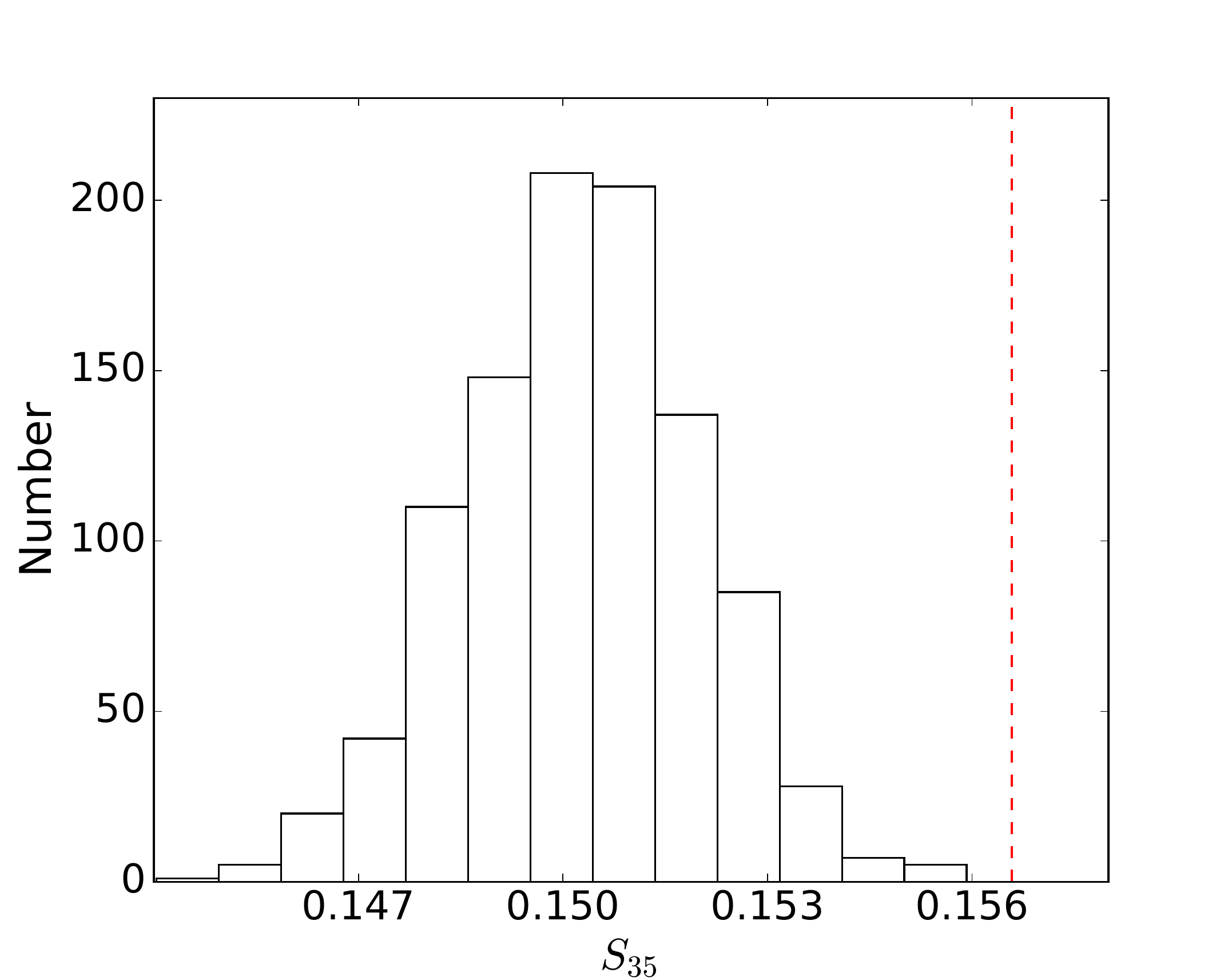}
			\caption{The distribution of the statistics $S_{35}$ for the $1\,000$ shuffled catalogues of the RGZ sample as presented in Sec.~\ref{sec:RandomDatasets}. The dashed red line marks the highly significant observed value. }
			\label{fig:S35}
		\end{figure}

	Using Eq. \eqref{eq:Maha} as a statistical test, we obtain $d^2 = 26.15$ for the RGZ sample, corresponding to a p-value $< 0.02$. On the plotted scales this signal is found not to be consistent with the noise. The distribution of $S_{35}$ for the shuffled catalogues (see Sec.~\ref{sec:RandomDatasets}) is plotted in Fig.~\ref{fig:S35}, together with the measured value.
	
	For the other two samples in Fig.~\ref{fig:Sn}, the signal is confirmed to be consistent with the noise (p-value $> 0.05$).
	
\begin{figure}
			\centering
			\includegraphics[width=.45\textwidth]{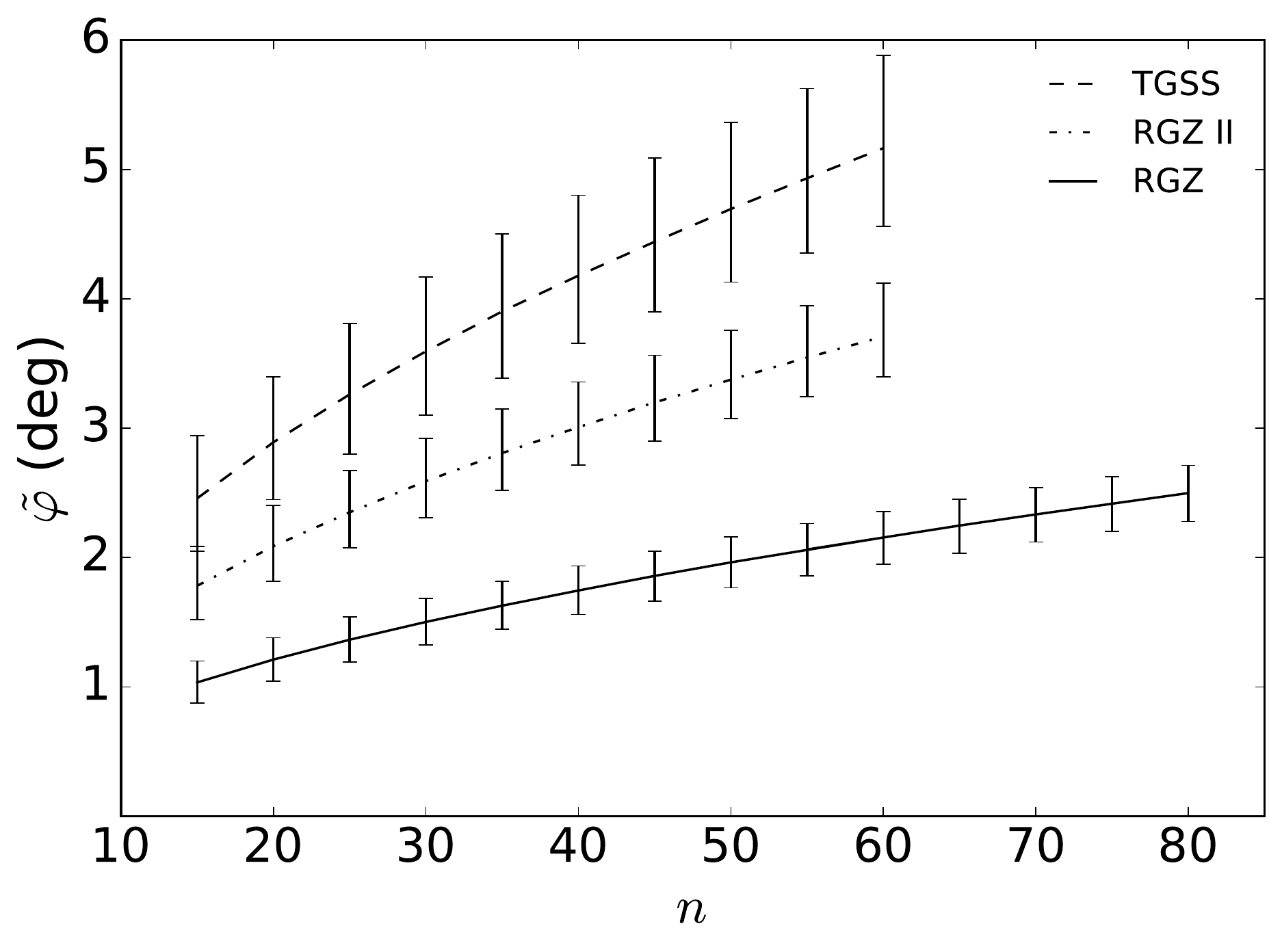}
			\caption{Median of the aperture radii probed by considering the $n$ closest neighbours as a function of $n$. The errorbars delimit the $16$th and $84$th percentile of the distributions. Two of three samples are described in Section~\ref{sec:SS} (TGSS and Radio Galaxy Zoo). The third, RGZ II, is a subsample of the RGZ sample designed to mimic the TGSS lower source density and source count.}
			\label{fig:sep}
		\end{figure}

\begin{figure}
			\centering
			\includegraphics[width=.45\textwidth]{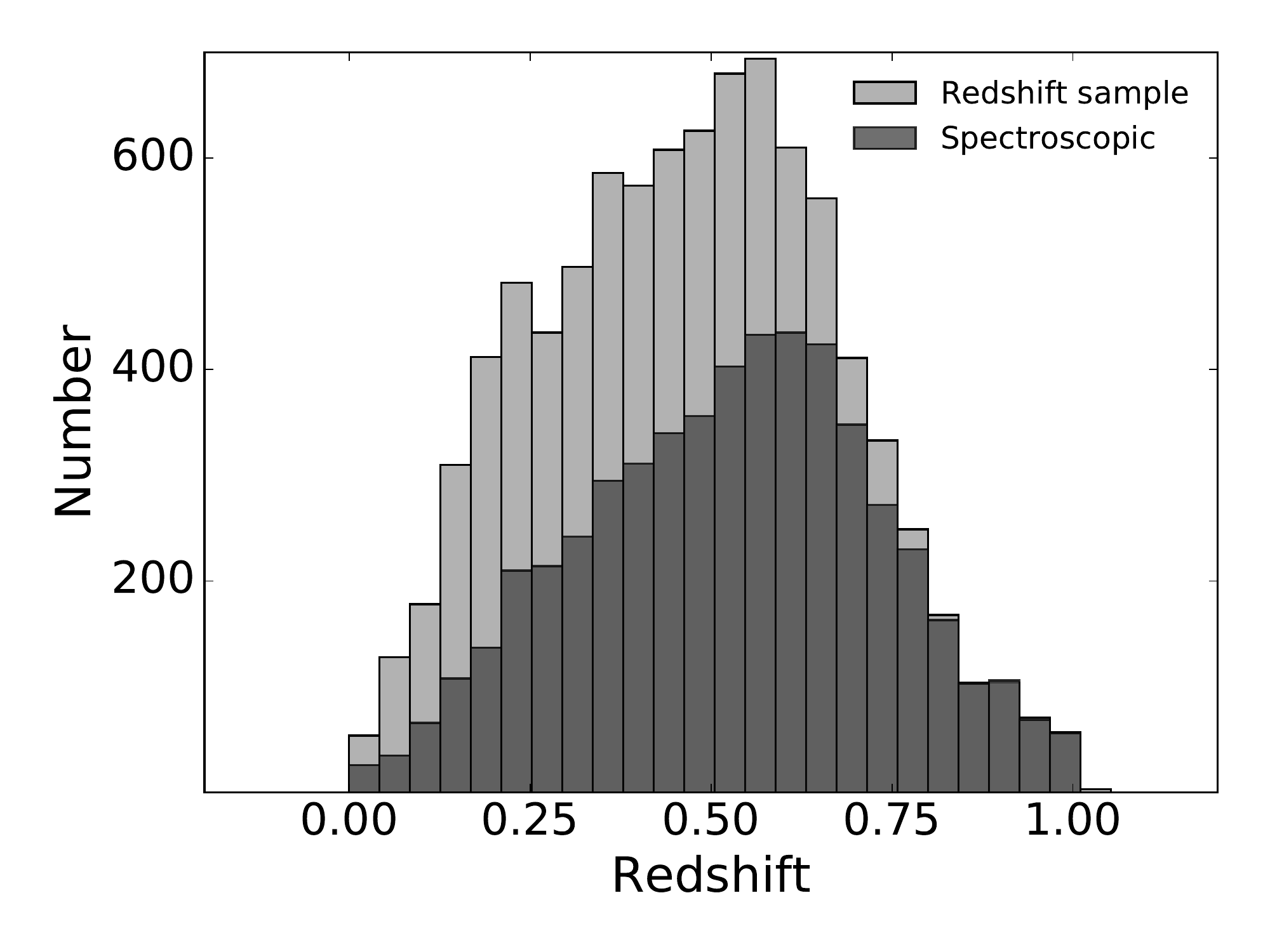}
			\caption{Redshift distribution of the selected sources in the Radio Galaxy Zoo sample. Around $13\%$ of the sources have photometric redshift and another $17\%$ of them have spectroscopic redshift. }
			\label{fig:z}
		\end{figure}

	The lower limit for the variable $n$ is set by the condition $n \gg 1$ and in our case we choose $n=15$. On the other hand, the upper limit can reach any value $n < N$, where $N$ is the total number of sources in the sample. For the maximum values of $n$, our choice was motivated by the corresponding angular scales. In Fig.~\ref{fig:sep} we plot the median value of the set of angular scales $\{\varphi_{i, n}\}$ probed as a function of every considered $n$, see Eq.~\eqref{eq:phii}. The errorbars delimit the $68\%$ interval centred on the median. For the RGZ sample the maximum $n=80$ corresponds to $\tilde{\varphi} \approx 2.5^\circ$. For $30\%$ of the sources in our sample, the Radio Galaxy Zoo consensus catalogue contains an optical counterpart with known redshift. Around two thirds of these are spectroscopic and the rest are photometric. Fig.~\ref{fig:z} presents the redshift distribution. The median value is $z = 0.47$ if we consider both classes, and $z = 0.54$ if we consider only spectroscopic redshifts. Assuming a flat $\Lambda$CDM Cosmology and cosmological parameters $\Omega_m=0.31, \Omega_\Lambda = 0.69$; the angular scale of $2.5^\circ$ is equivalent to a comoving scale of around $70-85$ $h_{70}^{-1}$ Mpc at these redshifts. This is the typical length of the longest low-redshift filaments of the cosmic web \citep{Tempel2014}. Since no redshift information is provided for the TGSS sample, we opt for a maximal $n$ corresponding to an angular scale of $\varphi = 5^\circ$.

	Of the two physical position angle samples considered, RGZ is the only one containing a signal significantly higher than the noise, consistently above the weak alignment threshold as defined in Section~\ref{sec:S}. Physically we would expect the alignment strength to decrease as a function of $n$. However, in Fig.~\ref{fig:Sn} we can see a minimum of the S.L. located between $n = 35$ and $n=40$, corresponding to an angular scale between $1.5^\circ$ and $2^\circ$ (Fig.~\ref{fig:sep}). This is due to the broader distribution of $d_{i, n}$ for small $n$, which lowers the significance of $S_n$. A similar effect is visible when the same statistic is employed elsewhere \citep[e.g.][]{Lamy2001}.
	
	We use the position of this minimum as an upper bound of the maximal alignment scale. To get an estimate of the physical scales probed by $n=40$, we then use the available redshift information (Fig.~\ref{fig:z}). For the $68\%$ redshift interval quoted in Table \ref{tab:surveys}, the angular size $\varphi = 1.5^\circ$ corresponds to transversal physical sizes in the range $[19, 38]$ Mpc.  These distances roughly correspond to differential redshifts along the line of sight of the order of $\Delta z \sim 0.01$.
	
	If the alignment signal is due to physical proximity we expect these to be the relevant scales. To validate physical proximity as a possible explanation, we confirm that, among the sources with known redshift, $\sim1.5\times10^3$ pairs have an angular separation within $1.5^\circ$ and redshift difference within $0.01$. Since only a third of the RGZ sample has known redshift, we can then estimate the number of physically close pairs as $3\times1.5\times10^3 = 4.5\times10^3$. Because of the large uncertainties on photometric redshifts, this value underestimates the number of real pairs.

	The absence of an alignment signal in TGSS is not surprising. When reduced to similar number densities and source counts the signal is not present in the RGZ sample either.	Number density and source count affect the final signal $S_n$ in different ways. 
	A lower number density has the effect of shifting the signal towards lower $n$. As visible in Fig.~\ref{fig:sep}, the maximum scale $\tilde{\varphi}$ probed with the RGZ sample for $n=80$ corresponds barely to the minimum scale probed with the RGZ II sample.
	
	At the same time, the number count does directly affect the chances of measuring a significant alignment, since the variance is dominated by the shot noise in Eq. \eqref{eq:sigmaest}. Evidently, a change of a factor $3$ in the number of sources $N$ is enough to erase the alignment signal.
	
	The alignment detection discussed above could be contaminated by large radio galaxies, whose lobes are aligned with each other, e.g., along the same position angle, but are counted as separate sources in the RGZ sample.  This can occur because the volunteers are only presented with a $3\arcmin \times 3\arcmin$ field centred on a FIRST catalogue position, so sources larger than that may go unrecognized.  As a rough check on the impact of this potential contamination, we examined the FIRST images of $35$ double-lobed radio galaxies, $3.5\arcmin$ to $10\arcmin$ in extent, drawn from a sample of $6000$  such sources $>1'$ in extent and with secure optical identifications, compiled by one of us \citep[HA, see e.g.,][]{Andernach2012}.  None of these sources appeared in our RGZ sample as two distinct sources.  We therefore conclude that the large source contamination is unlikely to be making a significant contribution, based on a) the low (undetected) probability of having both lobes in our sample, b) and the relative scarcity of large sources in general, ($\sim 3.5\%$ of FRII radio galaxies are $ 1.5\arcmin$, using figure 11 from \citep{Overzier2003}, and c) the fact that our highest significance signal occurs between $1.5$ and $2$ degrees, where there are only a handful of sources so large in the whole sky.  However, the existence of a small fractional population of sources that RGZ volunteers may not find should be investigated further when detailed size distributions are being studied.

\section{Conclusions}
	We constructed two samples of radio galaxies to search for the signature of source alignment: one based on the Radio Galaxy Zoo November 2015 catalogue, and the other on the TGSS Alternative Data Release 1 catalogue.
	
	The RGZ sample is formed by sources present in the FIRST survey and classified by volunteers participating in the Radio Galaxy Zoo collaboration. In this paper, we report marginal evidence of local alignment among radio sources within this sample. The signal is inconsistent with the noise  with a significance level $> 2\sigma$. Its main feature is a  $3.2\sigma$ minimum of the significance level on angular scales between $1.5^\circ$ and $2^\circ$. Assuming a flat $\Lambda$CDM Cosmology and cosmological parameters $\Omega_m=0.31, \Omega_\Lambda = 0.69$, this roughly corresponds to a physical scale in the range $[19, 38]$ Mpc.
	
	By number of sources, RGZ is about six hundred times larger than the set considered by \cite{Taylor2016} and about one hundred times larger than the largest set of quasars considered for the alignment study of quasar polarization vectors \citep{Pelgrims2014}. More detailed investigations of other, even larger samples, with different selection biases (see  Sec.~\ref{sec:SS}) or choices for the scales of interest (see Sec.~\ref{sec:RES}), would be useful.

	The TGSS sample was obtained from a reprocessed GMRT survey. In this case, no evidence of alignment is found. However, its lower source density means that even if a signal was present, it would not be significant. 
		
	The alignment of astronomical sources has frequently been a topic of interest. Optical galaxies have usually dominated the conversation \citep{Joachimi2015}, which in recent years has seen a resurgence in popularity due to the identification of galaxy alignment as a systematic effect for weak lensing \citep{Kirk2015}. If the alignment of radio galaxies is proved to be connected to the tidally induced alignment of their optical counterparts, radio observations might be used to constrain the intrinsic orientation of galaxies. 
	
	An alternative hypothesis might revolve around the origin of radio-loud AGNs, believed to be associated with galaxy mergers \citep[see, for example,][]{Hardcastle2007, Croton2006, Chiaberge2015}. If mergers play a role in spinning up the supermassive black hole or orienting the accretion disk emitting the jets, a preferential merger direction along the filaments of the large-scale structure could result in the alignment of the jets.
	
	With the new generation of high resolution radio interferometers like the Low Frequency Array (LOFAR) and the Square Kilometre Array (SKA), the cosmological prospects of radio astronomy will be expanded \citep[e.g.,][]{Blake2004a, VanHaarlem2013}. We expect the study of alignment to be part of these efforts.

\section*{Acknowledgements}

This publication has been made possible by the participation of more than 7000 volunteers in the Radio Galaxy Zoo project.  The data in this paper are the result of the efforts of
the Radio Galaxy Zoo volunteers.
Their efforts are individually acknowledged at {\url{http://rgzauthors.galaxyzoo.org}}.

This publication makes use of data product from the Karl G. Jansky Very Large Array. The National Radio Astronomy Observatory is a facility of the National Science Foundation operated under cooperative agreement by Associated Universities, Inc.

This publication makes use of data products from the Wide-field
Infrared Survey Explorer (WISE) and the Spitzer Space Telescope. The WISE is a joint project of the University of California, Los Angeles, and the Jet Propulsion Laboratory/California Institute of Technology, funded by the National Aeronautics and Space Administration. SWIRE is supported by NASA through the SIRTF Legacy Program under contract 1407 with the Jet Propulsion Laboratory. 

We also thank the staff of the GMRT that made possible the observations TGSS is based upon. GMRT is run by the National Centre for Radio Astrophysics of the Tata Institute of Fundamental Research.

FdG is supported by the VENI research programme with project number 1808, which is financed by the Netherlands Organisation for Scientific Research (NWO). Partial support for LR comes from US National Science Foundation grants AST-1211595 and AST-1714205 to the University of Minnesota. HA benefitted from grant DAIP 980/2016-2017 of the University of Guanajuato. Parts of this research were conducted by the Australian Research Council Centre of Excellence for All-sky Astrophysics (CAASTRO), through project number CE110001020.

%%%%%%%%%%%%%%%%%%%%%%%%%%%%%%%%%%%%%%%%%%%%%%%%%%


\bibliographystyle{mnras}
\bibliography{bibcars}

%%%%%%%%%%%%%%%%%%%%%%%%%%%%%%%%%%%%%%%%%%%%%%%%%%

%%%%%%%%%%%%%%%%% APPENDICES %%%%%%%%%%%%%%%%%%%%%

\appendix
	\section{Position Angle as Shear}
		\label{sec:shear}
		In this Appendix, we focus on an approach to the study of the position angles based on an alternative formalism. The study of other directional quantities over large scales through the use of spin-2 spherical harmonics is well established. Examples of such quantities are the polarization $P$ of the CMB or the cosmic shear field $\gamma$ \citep[e.g.,][]{Collaboration2015, Hikage2011}. However, in our attempts, the detailed properties of the position angle datasets forced a sampling of the correlation functions and power spectra that did not allow us to resolve features like the minimum in Fig.~\ref{fig:Sn}.  In particular, the main complications are the partial sky-coverage, the low source density and the predisposition to systematic effects of interferometric measurements.
		
		Although the products presented in this appendix are inconclusive, we describe here our implementation of the cosmic shear statistics, so that it can be applied when suitable samples will become available.
		
		Cosmic shear is usually detected through the analysis of the spin-2 field
		
		\begin{equation}
			\gamma = \gamma_1 + i\gamma_2,
			\label{eq:gamma}
		\end{equation}
		
		where $\gamma_1, \gamma_2$ are defined on a local Cartesian reference frame. Under rotation of an angle $\Phi$ the field transforms as $\gamma\to \gamma~e^{2i\Phi}$. The shear is usually estimated as the ensemble average of galaxy ellipticities $\varepsilon$ \citep{Kirk2015}
		
		\begin{equation}
			\varepsilon =  \frac{1-q}{1+q} (\cos 2\alpha_p + i \sin 2\alpha_p)
			\label{eq:epsi}
		\end{equation}
		\begin{equation}
			\gamma = \braket{\varepsilon}
		\end{equation}
		
		In this definition, $\alpha_p$ is the major axis position angle of the optical galaxy and $q$ is the ratio between the major and minor axes. We define the tangential and cross-component ellipticity $\varepsilon_t$ and $\varepsilon_\times$ with respect to a direction as the projection of the ellipticity in the two  $+/\times$ components: (1) parallel or perpendicular to it (2) oriented at $45^\circ$ or $-45^\circ$. For a direction defined by the polar angle $\Psi$
		
		\begin{gather}
			\epsilon_t = -\operatorname{Re}\{e^{-2i\Psi}\epsilon \}
			\label{eq:et}
			\\
			\epsilon_\times = -\operatorname{Im}\{e^{-2i\Psi}\epsilon\}
			\label{eq:ecross}
		\end{gather}
		
		In our sign convention, a positive $\varepsilon_t$ corresponds to tangential alignment, i.e., the position angle $\alpha_p$ and the direction $\Psi$ are parallel, while a negative value corresponds to radial alignment, i.e., the two are perpendicular \citep{Kilbinger2015}.

		The literature contains multiple statistics involving the shear field. In particular, we focus on those described in  \cite{Schneider2002}, \cite{Eifler2010} and implemented by the software \textsc{treecorr}\footnote{\url{https://github.com/rmjarvis/TreeCorr}} \citep{Jarvis2004}. 
		
		When evaluating a two point correlation function, the two components $\gamma_t$ and $\gamma_\times$ are defined with respect to the direction connecting the sources. These components are commonly estimated by neglecting both the curvature of the sphere and the parallel transport operation described in Section~\ref{sec:PT}. Because of this, we limit our analysis in this Section to distances smaller than $5^\circ$, corresponding to about $0.1$ radians.
		
		We introduce the two-point correlation functions
		
		\begin{gather}
			\xi_{tt}(\varphi) = \braket{\gamma_t \gamma_t}
			\label{eq:xitt}
			\\
			\xi_{\times \times}(\varphi) = \braket{\gamma_\times \gamma_\times}
			\label{eq:xicc}
			\\
			\xi_+ (\varphi)= \braket{\gamma_t \gamma_t} + \braket{\gamma_\times \gamma_\times} 
			\\
			\xi_-  (\varphi)= \braket{\gamma_t \gamma_t} - \braket{\gamma_\times \gamma_\times}
		\end{gather}
		
		where the averages are computed over every possible pair of sources with angular distance $\varphi$. The tangential and cross-component shear are defined as in Eq. \eqref{eq:et}, \eqref{eq:ecross}. The two correlation functions $\xi_{tt}$ and $\xi_{\times \times}$ distinguish between different shear configurations, according to the provided definitions of $\gamma_t$ and $\gamma_\times$.
		Furthermore, we define $\overline{\gamma}(\varphi)$ as the mean shear inside a circular aperture of radius $\varphi$. The variance of this quantity can then be estimated directly from the correlation function $\xi_+$
		
		\begin{gather}
			\braket{|\overline{\gamma}|^2}(\varphi) = \int \frac{d\vartheta\vartheta}{2\varphi^2} 
				\xi_+(\vartheta) S_+ \left( \frac{\vartheta}{\varphi}\right)
			\label{eq:Gsq}
		\end{gather}
		
		The definition of the weight function $S_+$ and a more detailed introduction to the top-hat shear dispersion are given by \cite{Schneider2002}.

		Using the representation introduced in Eq. \eqref{eq:epsi}, the position angle $\alpha$ can be written as
		
		\begin{equation}
			\gamma^\alpha = \cos 2\alpha + i \sin 2 \alpha
		\end{equation} 			
		
		Under a rotation of an angle $\Phi$ the quantity $\gamma^\alpha$ behaves exactly like the shear field, $\gamma^\alpha \to \gamma^\alpha~e^{2i\Phi}$. This justifies the extension to $\gamma^\alpha$ of the statistics defined for $\gamma$. Since we want to study the alignment configuration of the position angles, we should point out that no averaging is involved. In our analysis $\gamma^\alpha$ takes the place of the shear field $\gamma$ and not of the ellipticity $\varepsilon$.    
		
		In the presence of a global systematic effect we rewrite the correlation functions \eqref{eq:xitt} and \eqref{eq:xicc} as
		
		\begin{gather}
			\xi_{tt} (\theta) = \braket{\gamma^\alpha_t \gamma^\alpha_t} - \xi^n_{tt}
			\\
			\xi_{\times \times} (\theta) = \braket{\gamma^\alpha_\times \gamma^\alpha_\times}
			-
			\xi^n_{\times \times},
		\end{gather}
		
		where we subtracted a noise bias, to be estimated through simulated random data sets containing only the noise. The expression for the estimator \eqref{eq:Gsq} must be computed from these unbiased correlation functions.
		
		We do not assume any particular model for our analysis and we set as our primary objective the detection of a positive correlation. In its absence we expect the two-point correlation functions and the dispersion to be consistent with the noise on every scale $\varphi$.
		
		The function $\braket{|\overline{\gamma^\alpha}|^2}(\varphi)$ is closely related to $S_n$ (Eq. \eqref{eq:S}) since both of them estimate the average dispersion (or dispersion squared) of the position angles. The first one considers spherical caps of constant aperture radius $\varphi$, while the second considers caps with a constant number of sources $n$. The dispersion $\braket{|\overline{\gamma^\alpha}|^2}(\varphi)$ has the advantage of probing precise angular scales, but for non-uniformly distributed samples its value can be easily skewed by the sources in low density regions. Another drawback, due to our chosen implementation, is the lack of parallel transport in its computation.

		\subsection{Products}
		
\begin{figure}
			\includegraphics[width=0.45\textwidth]{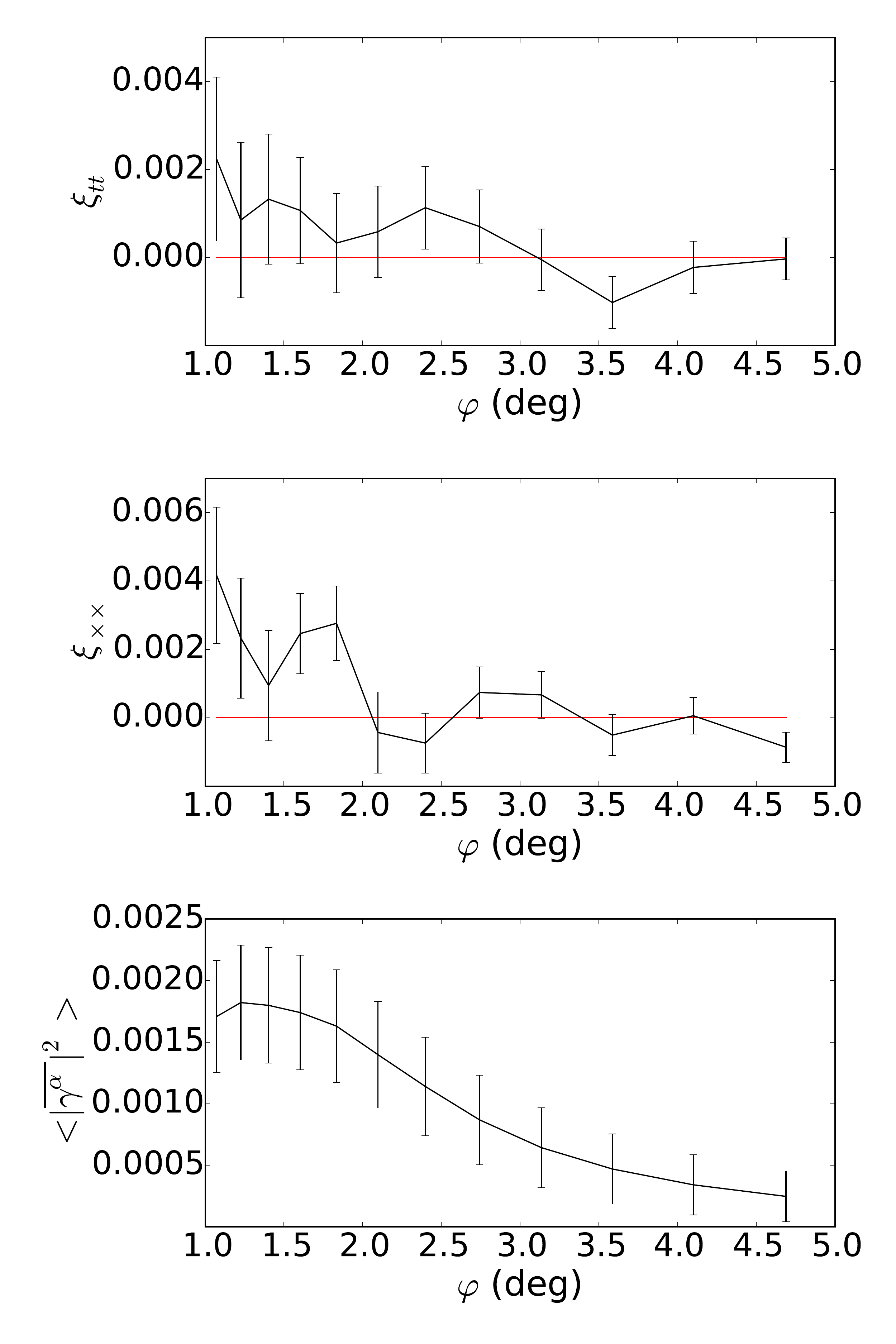}
			\caption{Weak lensing statistics for the Radio Galaxy Zoo sample: the two point correlation functions $\xi_{tt}(\varphi)$, $\xi_{\times \times}(\varphi)$ as a function of the distance $\varphi$ and the top-hat shear dispersion $\braket{|\overline{\gamma^\alpha}|^2}(\varphi)$ as a function of the aperture radius $\varphi$.}
			
			\label{fig:RGZweak}
		\end{figure}

\begin{figure}
			\includegraphics[width=0.45\textwidth]{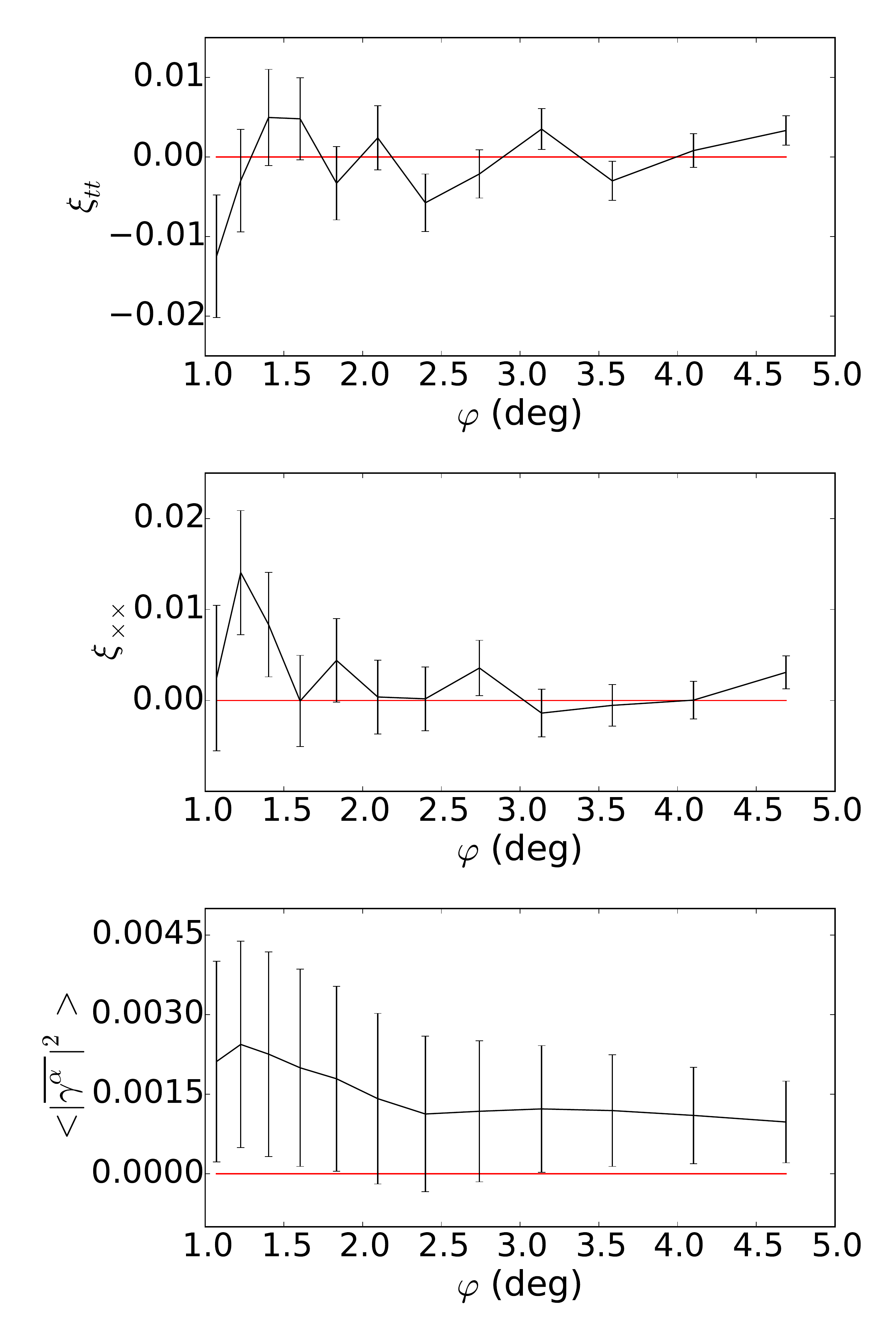}
			\caption{Weak lensing statistics for the TGSS sample: the two point correlation functions $\xi_{tt}(\varphi)$, $\xi_{\times \times}(\varphi)$ as a function of the distance $\varphi$ and the top-hat shear dispersion $\braket{|\overline{\gamma^\alpha}|^2}(\varphi)$ as a function of the aperture radius $\varphi$.}
			
			\label{fig:TGSSweak}
		\end{figure}

		In Fig.~\ref{fig:RGZweak} and~\ref{fig:TGSSweak} we plot the statistics presented in Eq. \eqref{eq:xitt}, \eqref{eq:xicc} and \eqref{eq:Gsq} for the Radio Galaxy Zoo and TGSS samples. The noise bias has already been subtracted. The covariance matrices are generated using the method described in Section~\ref{sec:RandomDatasets}.
		
		Since the diagonal terms in the covariance matrix \eqref{eq:cov} are two orders of magnitude higher than the non-diagonal terms, we can confirm that the measurements of the statistics $\xi_{tt}$ and $\xi_{\times \times}$ for different angular scales are in fact independent. The same is not true for the dispersion $\braket{|\overline{\gamma^\alpha}|^2}$. The reason for this is the same as the one discussed in Section~\ref{sec:S} for the statistics $S_n$.

		The correlation functions $\xi_{tt}(\theta)$ and $\xi_{\times \times}(\theta)$ are consistent with normally distributed noise. This result was checked using common statistical tests: (1) Shapiro-Wilk (2) $\chi^2$ (3) Anderson-Darling (4) two-tailed Kolmogorov-Smirnov. All of them returned p-values $> 0.05$. For the two $\braket{|\overline{\gamma^\alpha}|^2}$ we obtain the Mahalanobis distances $d^2 = 17.28$ and $d^2 = 12.05$. Given the number of degrees of freedom ($k=12$), both correspond to p-values $>0.05$, meaning that these results are also consistent with the noise. 
		
		Nothing conclusive about the alignment configuration can be stated, since both $\xi_{tt}$ and $\xi_{\times \times}$ are consistent with zero. The dispersion $\braket{|\overline{\gamma^\alpha}|^2}$ is also found to be consistent with the noise. This is not unexpected, since the estimator in Eq. \eqref{eq:Gsq} is simply a convolution of $\xi_{+} = \xi_{tt} + \xi_{\times \times}$ and a weight function. If $\xi_{+}$ is found to be largely consistent with zero, the same should be true for $\braket{|\overline{\gamma^\alpha}|^2}$.

			Finally, the down-crossing of $\xi_{tt}$ around the angular scale of $3\deg$ seems to suggest a change in the configuration of the alignment. The limited number of data points and the overall consistency with zero of the correlation function do not allow for a conclusive statement. However, assuming the downcrossing to be a feature, we can assign a significance to this observation. The probability of obtaining $8$ consecutive positive datapoints is found to be less than $0.005$. 


%%%%%%%%%%%%%%%%%%%%%%%%%%%%%%%%%%%%%%%%%%%%%%%%%%

% Don't change these lines
\bsp	% typesetting comment
\label{lastpage}
\end{document}